%% file: CVCS_GKP_arxiv.tex
\documentclass[aps,pra,twocolumn,tightenlines%
,amsmath,amssymb,amsfonts,nofootinbib,notitlepage%
,showpacs]{revtex4-1}

\usepackage{nag}
\usepackage{amsthm}
\usepackage{graphicx}
\usepackage{color}
\usepackage{bm}
\usepackage{mathtools}
\usepackage{qcircuit}
\usepackage[colorlinks=false, linkcolor=blue, citecolor=magenta]{hyperref}
\usepackage{mdwlist}

\theoremstyle{plain}

\theoremstyle{remark}

\DeclareMathOperator{\tr}{tr}

\def\tp{\mathrm{T}}

\def\logical{L}

\renewcommand{\Qcircuit}[1][0em]{\xymatrix @*[o] @*=<#1>}
\newcommand{\node}[2][]{{\begin{array}{c} \ _{#1}\  \\ {#2} \\ \
\end{array}}\drop\frm{o} }
\newcommand{\link}[2]{\ar @{-} [#1,#2]}

\def\clap#1{\hbox to 0pt{\hss#1\hss}}

\newcommand{\FT}{\text{FT}}
\newcommand{\err}{\text{err}}
\newcommand{\ideal}{\text{ideal}}
\newcommand{\success}{\text{succ}}
\newcommand{\corr}{\text{c}}
\newcommand{\botlab}{\text{bot}}
\newcommand{\toplab}{\text{top}}
\newcommand{\cool}{M}

\newcommand{\cG}{\mathcal G}
\newcommand{\cM}{\mathcal M}

\DeclareMathOperator\erf{erf}

\newcommand{\meas}{m}
\newcommand{\measvec}{\vec \meas}

\makeatletter
\@ifundefined{textcolor}{}
{%
 \definecolor{BLACK}{gray}{0}
 \definecolor{WHITE}{gray}{1}
 \definecolor{RED}{rgb}{1,0,0}
 \definecolor{GREEN}{rgb}{0,.4,0}
 \definecolor{BLUE}{rgb}{0,0,1}
 \definecolor{CYAN}{cmyk}{1,0,0,0}
 \definecolor{MAGENTA}{cmyk}{0,1,0,0}
 \definecolor{YELLOW}{cmyk}{.2,.4,1,0}
 }
\makeatother

\newcommand{\ra}[1]{\renewcommand{\arraystretch}{#1}}

\graphicspath{{graphics/}}

\def\negspace{\!}
\def\lsub#1#2{{\protect\vphantom{#1}}_{#2} \negspace {#1}}
\def\rsub#1#2{{#1} \negspace {\protect\vphantom{#1}}_{#2}}
\def\lrsub#1#2#3{{\protect\vphantom{#1}}_{#2} \negspace {#1} \negspace {\protect\vphantom{#1}}_{#3}}

\def\ketsub#1#2{\rsub {\ket{#1}} {#2}}
\def\brasub#1#2{\lsub {\bra{#1}} {#2}}

\def\qbra#1{\brasub{#1} q}

\def\qket#1{\ketsub{#1} q}

\def\inprod#1#2{\left\langle {#1} | {#2} \right\rangle}

\def\inprodsubsub#1#2#3#4{\lrsub {\inprod{#1}{#2}} {#3} {#4}}

\def\outprod#1#2{\ket {#1}\!\bra {#2}}

\def\avg#1{\left\langle {#1} \right\rangle}

\def\abs#1{\left\lvert{#1}\right\rvert}

\def\wholenums{\mathbb{N}_0}
\def\integers{\mathbb{Z}}

\def\op#1{\hat{#1}}
\def\opvec#1{\op{\vec{#1}}}

\def\id{I}

\def\1{\mat{\id}}

\def\mat#1{\bm{\mathrm{#1}}}

\renewcommand{\vec}[1]{\bm{\mathrm{#1}}}

\def\controlled#1{\op{\mathrm{C}}_{#1}}
\def\CZ{\controlled Z}

\newcommand{\sectionPRL}[1]{\emph{{#1}.}---}
\newcommand{\acknowledgementsPRL}{\sectionPRL{Acknolwedgments}}

\begin{document}

\title{Fault-Tolerant Measurement-Based Quantum Computing \\ with Continuous-Variable Cluster States}

\author{Nicolas C. Menicucci}
\affiliation{School of Physics, The University of Sydney, Sydney, NSW, 2006, Australia}
\email{ncmenicucci@gmail.com}

\date{\today}

\begin{abstract} 
A long-standing open question about Gaussian continuous-variable cluster states is whether they enable fault-tolerant measurement-based quantum computation. The answer is yes. 
Initial squeezing in the cluster above a threshold value of 20.5~dB ensures that errors from finite squeezing acting on encoded qubits are below the fault-tolerance threshold of known qubit-based error-correcting codes. By concatenating with one of these codes and using ancilla-based error correction, fault-tolerant measurement-based quantum computation of theoretically indefinite length is possible with finitely squeezed cluster states.
\end{abstract}


\maketitle

\sectionPRL{Gaussian cluster states}%
Quantum computing~(QC) harnesses inherently nonclassical features of quantum physics to perform computations that would be impractical for any ordinary (classical) computer~\cite{Nielsen2000}. This requires making quantum systems interact in a carefully controlled and coherent manner, which is often very difficult. On the other hand, measuring quantum systems is usually much easier. Measurement-based QC makes use of this fact, replacing %
 the difficulty of coherently controlling interactions between quantum systems with the up-front challenge of creating an entangled resource known as a cluster state~\cite{Briegel2001}, whereafter local adaptive measurements alone enable the full power of QC~\cite{Raussendorf2001}.
 
Normally in a quantum computer, quantum information is stored in qubits~\cite{Nielsen2000}, but continuous-variable~(CV) approaches also exist~\cite{Lloyd1999} in which wavefunctions over a continuous quantum variable are the basic information carriers. When it comes to measurement-based QC, optical CV cluster states~\cite{Zhang2006,Menicucci2006} offer a distinct advantage over their optical-qubit counterparts~\cite{Yao:2012fp,Su:2012jo} because they are much easier to make experimentally~\cite{Yokoyama:2013jp,Pysher:2011hn,Chen:2013tw}. In fact, highly scalable experimental designs exist for creating very large CV cluster states~\cite{Wang:2013tl,Menicucci2011a,Menicucci2010,Menicucci2008,vanLoock2007}, and an experimentally demonstrated 10,000-mode CV cluster state~\cite{Yokoyama:2013jp} now holds the world record for the largest entangled state ever created in which each constituent quantum system (in this case, a temporal packet of light) is individually addressable. This shatters the previous record of 14 trapped ions~\cite{Monz2011} by three orders of magnitude. Even more recently, a frequency-encoded CV cluster state has claimed second place with 60 entangled frequency modes and the promise of thousands more available~\cite{Chen:2013tw}.

This ease of experimental generation and scalability comes at the price of inescapable noise when these states are used for quantum information processing~\cite{Gu2009,Alexander:MGl69zxv}. Ideal CV cluster states are unphysical~\cite{Menicucci2011}, so when discussing their physical realization, one always speaks of Gaussian states~\cite{Weedbrook:2012fe} for which certain linear combinations of quadrature variables have reduced variance (i.e.,~squeezing)~\cite{Gu2009,Menicucci2011}. As these variances tend to zero, or equivalently the squeezing tends to infinity, these states become better and better approximations to ideal CV cluster states~\cite{Menicucci2011}, but the required energy diverges. Keeping the energy finite requires that the squeezing remain finite, which means that even with perfect experimental equipment, information degradation is inevitable when using CV cluster states for measurement-based QC.

When used in the real world, both qubit and CV cluster states will suffer from noise, but one might wonder whether the \emph{intrinsic noise} of CV cluster states due to finite squeezing might be fundamentally different in some way. Previous results showed that there is no easy fix for this type of noise~\cite{Ohliger2010,Ohliger2012} and left hanging in the air the question of whether finitely squeezed (and thus physical) CV cluster states were at all useful for practical measurement-based QC of indefinite length. If not, it would mean there were a fundamental deficiency in CV cluster states not suffered by their qubit-based cousins. The possibility remained, however, that the noise might be handled using well-established methods of error correction and fault tolerance~\cite{Shor:1996jj,Aharonov:1997kc,Knill:1998bv,Preskill:1998gf,Kitaev:1997kq,Knill2005,Aliferis:2007wb,Gottesman1997} applied to qubits encoded as CV wavefunctions (e.g.,~\cite{Gottesman2001}), a possibility that the authors themselves point out~\cite{Ohliger2010}.

Fault-tolerant QC (see Ref.~\cite{Gottesman:2009ug} for a review) is the ability to reduce logical errors in a quantum computation to arbitrarily low levels if the physical error rate of the individual gates comprising the computation is below a fixed, positive value called the \emph{fault-tolerance threshold}. In other words, if the probability of error in every physical gate can be guaranteed to be below the threshold, then these noisy gates can be used to implement quantum error correction of noisy quantum information in a way that can make the computation's overall error rate as low as one desires, no matter how long the computation.

Qubit cluster states admit a fault-tolerance threshold for measurement-based QC~\cite{Nielsen:2005fd,Aliferis2006}, which can be made strikingly high~(1.4\%) using topological methods~\cite{Raussendorf2006} and which can be further refined to a few percent by postselection~\cite{Silva:2007ey,Fujii:2010fe,Fujii:2010kx}. Fault-tolerance thresholds for more traditional codes (i.e.,~concatenated codes) vary, with typical thresholds being $10^{-6}$~\cite{Preskill:1998gf,Kitaev:1997kq,Aharonov:1997kc,Knill:1998bv}, $10^{-4}$~\cite{Gottesman1997}, and $3\times 10^{-3}$~\cite{Steane:2003gp}---and up to a few percent with postselection~\cite{Knill2005}.

Since one can, in principle, implement any unitary on CV-encoded quantum information using a CV cluster state (albeit noisily), our strategy will be to encode qubits as CV wavefunctions~\cite{Gottesman2001} in a way that maps the natural noise of a CV cluster state into noise on the gates processing the encoded qubits. Higher squeezing will produce a lower gate error rate. If the squeezing is high enough, this error rate will be below the threshold for some known error-correcting code as discussed above, and we can use the CV cluster state to implement fault-tolerant QC on the encoded qubits. 
Our goal, then, will be to prove the existence of a \emph{squeezing threshold}: a constant, finite level of squeezing above which fault-tolerant measurement-based QC is possible using encoded qubits and a concatenated error-correcting code, assuming no other noise beyond that introduced by finite squeezing alone~\cite{Gu2009,Alexander:MGl69zxv}.

\sectionPRL{GKP-encoded qubits and Gaussian channels}%
The qubit encoding of Gottesman, Kitaev, and Preskill~(GKP)~\cite{Gottesman2001} in its simplest form encodes one qubit per oscillator. The position-space wavefunction for each of the logical computational basis states is an evenly spaced comb of $\delta$-functions separated by~$2\sqrt\pi$, and the two states' combs are offset by $\sqrt \pi$ from each other---specifically, $\ket{j_\logical} \propto \sum_{s \in \integers} \qket {(2s+j) \sqrt\pi}$ ($j=0,1$), where $\qket s$ is an eigenstate of position for the oscillator. A physical realization of this encoding replaces the $\delta$-functions with sharp Gaussians and limits their heights according to a large Gaussian envelope. Although challenging to create, proposals exist to generate such states optically~\cite{Vasconcelos:2010gb} or by a variety of other methods~\cite{Gottesman2001,Travaglione:2002fp,Pirandola:2004jo,Pirandola:2006gh,Pirandola:2006bf}.

This encoding protects quantum information against random shifts in the quadrature variables $\op q$~(position) and $\op p$~(momentum)~\cite{Gottesman2001}. When Gaussian distributed, a random shift is called a Gaussian channel and can be modeled as Gaussian convolution of the input Wigner function. This is exactly the noise model of CV cluster-state QC~\cite{Gu2009,Alexander:MGl69zxv}, making GKP an appealing qubit encoding---as long as error correction can be performed with minimal deviation from the measurement-based paradigm (Cf. Ref.~\cite{Aliferis2006}). Fortunately, GKP error correction~\cite{Gottesman2001} dovetails nicely with CV cluster states, with details found in Appendix~\ref{subapp:GKPEC}. 

\sectionPRL{Fault-tolerant Clifford gates}%
The workhorse of (qubit-based) fault-tolerant quantum computation is the Clifford group~\cite{Nielsen2000,Gottesman1997}, which can be generated by supplementing the Pauli group with the single-qubit gates of Hadamard and phase, as well as with a two-qubit gate such as the controlled-$Z$ gate (sometimes called CPHASE). We need to be able to perform all of these gates with a below-threshold error rate.

The GKP-encoded Pauli group is just the CV Weyl-Heisenberg group restricted to shifts by integer multiples of $\sqrt \pi$ in position and/or momentum~\cite{Gottesman2001}. In CV measurement-based quantum computation, such displacements are ubiquitous and are therefore considered free to implement, and everything else is done with measurements~\cite{Gu2009}. GKP-encoded Hadamard and phase gates correspond to the Fourier transform~$\op F = e^{i\frac \pi 4(\op q^2 + \op p^2)}$ and shear~$\op P = e^{\frac i 2 \op q^2}$, respectively, and the qubit controlled-$Z$ gate is just a CV $\CZ$ gate with weight~$\pm 1$ ($\CZ[\pm 1] = e^{\pm i\op q \otimes \op q}$)~\cite{Gottesman2001}. All of these CV operations are Gaussian unitaries, which are easy to implement on a CV cluster state~\cite{Gu2009}. This is a huge advantage because it means the entirety of the GKP-encoded Clifford group inherits this ease of implementation.

Any single-mode Gaussian unitary can be implemented using four quadrature measurements, $\{\op p + \meas_j \op q\}_{j=1}^4$, on a linear CV cluster state (a.k.a.~CV quantum wire)~\cite{Ukai:2010bi}. %
We define the \emph{measurement vector} $\measvec = (\meas_1, \dotsc, \meas_4)$ to be the vector containing the four \emph{shearing parameters}~\cite{Alexander:MGl69zxv} associated with the quadrature measurements. $\measvec^{(\id)} = (0,0,0,0)$, $\measvec^{(F)} = (1,1,1,0)$, and $\measvec^{(P)} = (1,0,0,0)$ implement the identity~$\op \id$, Fourier transform~$\op F$, and shear~$\op P$, respectively. 
The following piece of an ancilla-supplemented CV cluster state allows these Gaussian unitaries to be implemented on the input state~$\ket\phi$, followed by GKP error correction (blank nodes are $\op p$-squeezed vacuum states; nodes with $0_\logical$ are GKP-encoded ancillas~$\ket {0_\logical}$; links are $\CZ$ gates with weight~$+1$~\cite{Gu2009}):
\begin{equation}
\label{eq:singleGaussoncluster}
    \Qcircuit[1.6em] @R=0.75em @C=0.75em {
    & & & \node{0_\logical} \link{1}{0} & \node{0_\logical} \link{1}{0} \\
    \node{\phi} {} & \node{} \link{0}{-1} & \node{} \link{0}{-1} & \node{} \link{0}{-1} & \node{} \link{0}{-1}  \\
    \text1 & 2 & 3 & 4 & \text{(out)}
    }
\end{equation}
To implement gate~$\op G$, we must perform quadrature measurements associated with $\measvec^{(G)}$ on nodes 1--4 on the bottom row. %
Measuring the ancillas (and appropriately displacing the nodes below) performs the GKP error correction on both $\op q$ and~$\op p$, as shown in Appendix~\ref{subapp:GKPEC}. %
To apply gates sequentially, one identifies the output node with node~1 of the next cluster.

Implementing a CV $\CZ$ gate requires links in the second lattice dimension~\cite{Gu2009}. Here is an ancilla-supplemented cluster that implements a GKP error-corrected $\CZ$ gate:
\begin{equation}
\label{eq:CVCZoncluster}
    \Qcircuit[1.6em] @R=0.75em @C=0.75em {
    &  &  & \node{0_\logical} \link{1}{0} & \node{0_\logical} \link{1}{0} 
    \\
    \node{\phi} {} & \node{} \link{0}{-1} & \node{} \link{0}{-1} & \node{} \link{0}{-1} & \node{} \link{0}{-1} & \text{(out)}
    \\
   \node{} \link{-1}{0} \link{1}{0} %
    \\
   \node{} \link{-1}{0} \link{1}{0}  & & & \node{0_\logical} \link{1}{0} & \node{0_\logical} \link{1}{0} 
    \\
   \node{\psi} {} & \node{} \link{0}{-1} & \node{} \link{0}{-1} & \node{} \link{0}{-1} & \node{} \link{0}{-1}  & \text{(out)}
    \\
    }
\end{equation}
Measuring every mode in~$\op p$ except the two %
output modes 
implements a $\CZ$~gate on the input state~$\ket\phi \otimes \ket\psi$, followed by two GKP error-corrected identity gates (see Appendix~\ref{subapp:CZ} for details). %
Single-mode gates [using Cluster~\eqref{eq:singleGaussoncluster}] and $\CZ$~gates [using Cluster~\eqref{eq:CVCZoncluster}] can be combined into an arbitrary Clifford circuit by identifying each output node with the input of the next gate and including additional identity gates where required. While this is undoubtedly not the most efficient implementation, it is the simplest for a proof-of-principle demonstration of fault tolerance, which is the goal of this work.

\sectionPRL{Concatenated codes}%
GKP error correction projects the Gaussian noise into a particular shift error using (slightly noisy) ancillas. This shift error is then corrected by shifting back to the codespace in the direction that corresponds to the shift being smallest. If the shift is too big, a qubit-level logical error results. The error rate is determined by the initial noise in the data register and in the ancilla~\cite{Glancy2006}. After error correction in both quadratures, however, the original data-register noise has been completely replaced by independent, uncorrelated noise from the ancillas, thereby converting the Gaussian noise (from propagation through the cluster) into \emph{local, independent Pauli errors after each gate}. Thus, noise correlations cannot build up between distant data registers.

By abstractly treating the GKP error-corrected gates as faulty qubit gates, we can concatenate the GKP error correction with a qubit-level error-correcting code~\cite{Gottesman:2009ug} and completely forget about the fact that, at the physical level, we are using CV information processing. Then, if the error rate is low enough (discussed next), we can implement Clifford gates fault tolerantly by further concatenation. At that point, the only other ingredient required is the ability to distill a ``magic state'' for use in implementing a non-Clifford gate (discussed subsequently).

\begin{figure}[!tbp]
\includegraphics[width=0.9\columnwidth]
{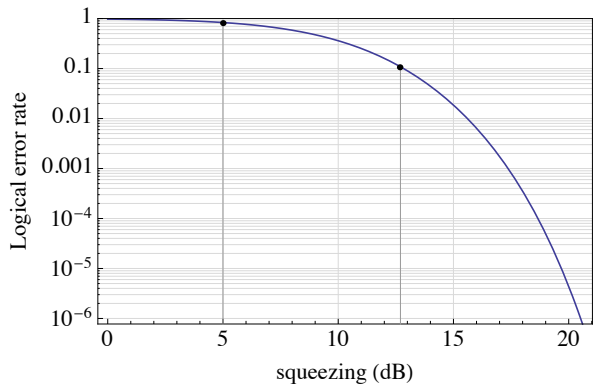}
\caption{Qubit-level logical error rate induced by GKP error correction with CV cluster states. The indicated level of squeezing is assumed to apply both to the initial momentum-squeezed states used to create the cluster state and in the Gaussian spikes that comprise the encoded GKP states. Also shown: maximum single-mode squeezing achieved to date (12.7~dB)~\cite{Eberle:2010jh,Mehmet:2011bu} and squeezing achieved in a large CV cluster state~(5~dB)~\cite{Yokoyama:2013jp}.%
}
\label{fig:errorrates}
\end{figure}
\sectionPRL{Squeezing threshold}%
To determine the amount of squeezing required for fault-tolerant QC, we use a physically motivated model of encoded states in which the Wigner function for an ideal GKP-encoded state, which is a regular lattice of $\pm$~$\delta$~functions~\cite{Gottesman2001}, is replaced by a corresponding lattice of sharp $\pm$~Gaussian spikes, each of which has the same ${2\times 2}$ covariance matrix~$\mat\eta$, which we call the \emph{error matrix}. For these states to have finite energy, we require that the height of a given Gaussian spike is itself distributed according to a (very large) Gaussian envelope in both quadratures. This is consistent with the original proposal by GKP but extended to the possibility of larger envelopes, which correspond to mixed states. 
Because $\mat\eta$ is the same for each spike, the height of each spike is irrelevant in measurements of $\op q \mod \sqrt \pi$, which are used for error correction, and we can focus on $\mat \eta$ alone. %

\begin{table}[!tbp]
\begin{tabular}{| l | c | c | c | c | c | c |}
\hline
threshold $p_\FT$					
& $10^{-1}$	& $10^{-2}$	& $10^{-3}$	& $10^{-4}$	& $10^{-5}$	& $10^{-6}$ 	\\ \hline
variance $\sigma^2$ ($\times 10^{-3}$)	
& 26.0		& 13.8		& 9.16		& 6.80 		& 5.38		& 4.44 		\\ \hline
squeezing $s$~(dB)
& 12.8		& 15.6		& 17.4		& 18.7		& 19.7		& 20.5		\\ \hline
\end{tabular}
\caption{Squeezing required to achieve a given fault-tolerant error threshold $p_\FT$ using GKP-encoded qubits, expressed in~dB and as a variance (vacuum variance~$= \tfrac 1 2$). This level of squeezing is required in both the initial momentum-squeezed states used to create the cluster state and in the Gaussian spikes that comprise the encoded GKP states.
}
\label{tab:thresholds}
\end{table}

Specializing the method of Ref.~\cite{Glancy2006} to Gaussian-distributed shifts, we establish a minimum squeezing threshold as follows. Consider that the GKP encoding can perfectly correct a shift error when the magnitude of the shift error, plus the magnitude of the error in the ancilla used to measure the shift, is less than $\sqrt\pi/2$~\cite{Gottesman2001,Glancy2006}. %
When this bound is exceeded, a qubit-level logical Pauli error occurs because the state is ``shifted back'' in the wrong direction. Also note that there are two corrections ($\op q$ and~$\op p$) per mode per gate. For the gate to be free of error, \emph{all} of these corrections must succeed.

The calculation proceeds, then, by identifying which gate has the largest probability of logical error as the error matrix evolves through Cluster~\eqref{eq:singleGaussoncluster} using measurement vectors $\measvec^{(\id)}$, $\measvec^{(F)}$, and $\measvec^{(P)}$ and through Cluster~\eqref{eq:CVCZoncluster} using just $\op p$~measurements. %
Appendix~\ref{app:clifford} contains the details of the calculation; here we simply present the results.

The noisiest gate is the $\CZ$ gate, so it sets the noise threshold. There are four corrections in this case. Assuming that the initial variance~$\sigma^2$ in the Gaussian spikes of the encoded ancillas is the same as that of the initial momentum-squeezed vacuum states used to make the CV cluster state, two of the Gaussian-distributed shift errors (including ancilla noise) have variance~$7\sigma^2$, and two others have variance~$5\sigma^2$. Therefore, the probability that at least one of those corrections fails is
\begin{align}
\label{eq:perrfull}
	p_{\text{err}} = 1 - \left[\erf \left( \frac {\sqrt \pi} {2 \sqrt {14}\sigma} \right)\right]^2 \left[\erf \left( \frac {\sqrt \pi} {2 \sqrt {10}\sigma} \right)\right]^2 \,.
\end{align}
When $p_{\text{err}} < p_\FT$ for the fault-tolerance threshold~$p_\FT$ for some qubit error-correcting code~\cite{Gottesman:2009ug}, we can concatenate the GKP code with that code and perform fault-tolerant measurement-based quantum computation. The variance~$\sigma^2$ identified by this condition corresponds to a squeezing threshold of
\begin{align}
	s > -10 \log_{10} \left(\frac {\sigma^2} {1/2} \right)\,.
\end{align}
For $p_\FT = 10^{-6}$, which is a typical (and rather conservative) threshold for concatenated codes~\cite{Preskill:1998gf,Aharonov:1997kc,Knill:1998bv}, this means that $\sigma^2 < 4.44 \times 10^{-3}$, which corresponds to $s > 20.5$~dB. Figure~\ref{fig:errorrates} shows a plot of this curve for intermediate squeezing levels, while Table~\ref{tab:thresholds} lists the squeezing corresponding to several other typical threshold values. %

\sectionPRL{Magic-state distillation}%
With nearly perfect Clifford gates in hand, computational universality is achieved by guaranteeing the ability to distill a so-called \emph{magic state} from many noisy copies~\cite{Bravyi:2005dx}. The procedure doesn't have to work every time, but when it does work, it has to produce a noisy state with sufficient fidelity to the state of interest. Fortunately, the noise thresholds for magic-state distillation are as high as \text{14--17\%} in some cases~\cite{Reichardt:2005er}, significantly less stringent than the Clifford-gate requirements of $\sim 10^{-6}$. As such, we can effectively ignore the errors introduced by the Clifford operations entirely~\cite{Bravyi:2005dx,JochymOConnor:2013wb,Brooks:2013wt}.

Previous work has focussed on the cubic phase state~\cite{Gottesman2001,Gu2009}, but distilling this state requires an asymmetric noise model~\cite{Gottesman2001}. The natural noise model of CV cluster-state QC is symmetric in $\op q$ and $\op p$ on average~\cite{Gu2009,Alexander:MGl69zxv}, which is preferred when distilling an encoded Hadamard eigenstate~$\ket{\pm H_\logical}$~\cite{Gottesman2001}. Either state can be used to implement an encoded $\tfrac \pi 8$ gate~\cite{Gottesman2001}. %

Since $\op F \ket{\pm H_\logical} = \pm \ket{\pm H_\logical}$, a Hadamard eigenstate can be constructed by counting photons on one half of an encoded Bell pair~\cite{Gottesman2001}, which can be created by applying a $\CZ$ gate to $\ket {+_\logical} \otimes \ket {+_\logical}$, using measurements as discussed above. Then, we count photons on one side and obtain an outcome $n$. In the ideal case, if $n \mod 4 = \{0, 2\}$, then the remotely prepared state is $\ket{\pm H_\logical}$, respectively (and an odd $n$ is impossible). In the physical case, of course, errors in the encoded Bell pair will reduce this fidelity of identification and corrupt the average output state. As such, if we get an odd $n$, we know an error has occurred, so we discard the state and start over. If $n$ is even, then $\varepsilon$ is the probability that it reveals the wrong state at the output.

Ref.~\cite{Reichardt:2005er} identifies $\varepsilon<0.146$ as a tight threshold for being able to distill the resulting state~\cite{Bravyi:2005dx}, and this threshold holds even when distilling using noisy Clifford gates~\cite{Brooks:2013wt}%
. Assuming we begin with pure ancillas, the error probability~$\varepsilon$ is between $12.5\%$ and~$12.6\%$ for all squeezing values shown Table~\ref{tab:thresholds}, with a success probability (i.e., probability of obtaining an even outcome) of~${2/3}$. Since $\varepsilon < 14.6\%$, distillation is possible, thus completing the proof of fault tolerance for measurement-based QC using CV cluster states. Appendix~\ref{app:magic} contains the details of the calculation, as well as some possible ways to optimize this method. %

\sectionPRL{Universal resources}%
Since the clusters used to perform Clifford gates and distill magic states all fit within a regular square lattice, we can create a universal resource by starting with an ordinary square-lattice CV cluster state of sufficient size and attaching GKP ancillas at regular intervals, like flowers growing in a regular pattern in the ``garden'' of the original lattice. One can even measure the ancillas directly after attachment. Either way, attaching the ancillas early means we are using a non-Gaussian resource state, evading known no-go results~\cite{Ohliger2010,Ohliger2012}.

Alternatively, one can think of the act of attaching ancillas and measuring $\op p$ as a single operation of \emph{nondestructively} measuring $\op q \mod \sqrt \pi$ (with some noise). Thus, we can simply add to our toolbox of measurements a nondestructive measurement of $\op q \mod \sqrt\pi$ and view the original Gaussian cluster states as universal for fault-tolerant quantum computation using this augmented suite of measurements. This evades the no-go results of Refs.~\cite{Ohliger2010,Ohliger2012} because %
active error correction and concatenation are being used, which mean that the required size of the encoding will grow (albeit slowly) with the length of the computation~\cite{Aharonov:1997kc}.

\sectionPRL{Extensions}%
While this analysis focuses exclusively on finite-squeezing noise, it can be straightforwardly generalized to include additional local Gaussian noise, photon loss, and detector inefficiency. While these extensions will generalize the threshold to also be a function of the additional noise parameters, they are not expected to change the fundamental result, which is the existence of some finite threshold. %

\sectionPRL{Conclusion}%
This is a theoretical breakthrough in our understanding of what is possible using measurement-based quantum computation with continuous-variable cluster states. With an appropriate qubit encoding, active error correction, and initial squeezing above a constant finite threshold, continuous-variable cluster states are universal for fault-tolerant measurement-based quantum computation of indefinite length. %

While the encoding scheme presented here may be non-optimal due to the prohibitive nature of the required states, it has at least the flavor of practicality since multiqubit Clifford operations require only Gaussian unitaries. Furthermore, the existence of a finite squeezing threshold for continuous-variable cluster states when using this encoding may well spur new experimental developments in implementing these challenging states.

Regardless of the scheme's feasibility, a finite squeezing threshold is now known to exist for continuous-variable cluster states. This means that work can continue with confidence toward designing better schemes, improving the threshold, and achieving higher levels of squeezing%
.

\acknowledgementsPRL%
I am grateful to Steven Flammia, Stephen Bartlett, Akimasa Miyake, and Peter van~Loock for discussions. This work was supported by the Australian Research Council under grant No.~DE120102204.

\appendix
\input{CVCS_GKP_supp.tex}
\bibliographystyle{bibstyleNCM_papers}
\bibliography{allrefs}

\end{document}

%% file: CVCS_GKP_supp.tex
\section{Clifford gate error rates and squeezing thresholds}%
\label{app:clifford}

\subsection{GKP error correction using CV cluster states}%
\label{subapp:GKPEC}
The original circuits (Fig.~4 in Ref.~\cite{Gottesman2001}) for error correction using GKP states can be written in a form that fits nicely within the framework of CV cluster states. Here is an equivalent circuit for detecting a position shift~$s_q$ on some initial quantum data~$\ket \phi$:
\begin{equation}
\label{circ:qcorrect}
\includegraphics[width=0.4\columnwidth]{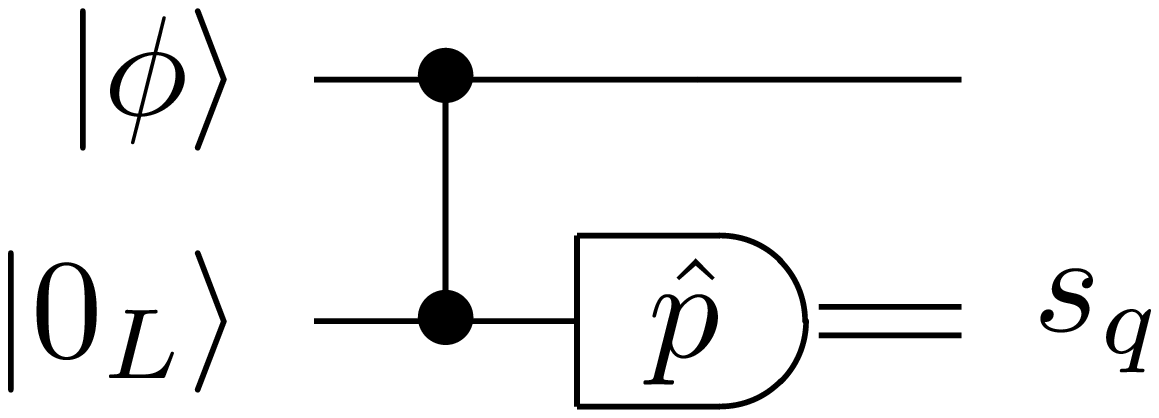}
\end{equation}
Applying $\op X[-(s_q \mod \sqrt \pi)]$, where the modulo function has range~$[-\sqrt\pi/2, \sqrt\pi/2)$, will correct the data, or we can simply note this change of basis in the rest of our computation. To correct momentum shifts, one can
sandwich Circuit~\eqref{circ:qcorrect} between Fourier and inverse Fourier transforms on the data.
Using two ancillas, then, we can correct for both types of error in sequence:
\begin{equation}
\label{circ:bothcorrect}
\includegraphics[width=0.5\columnwidth]{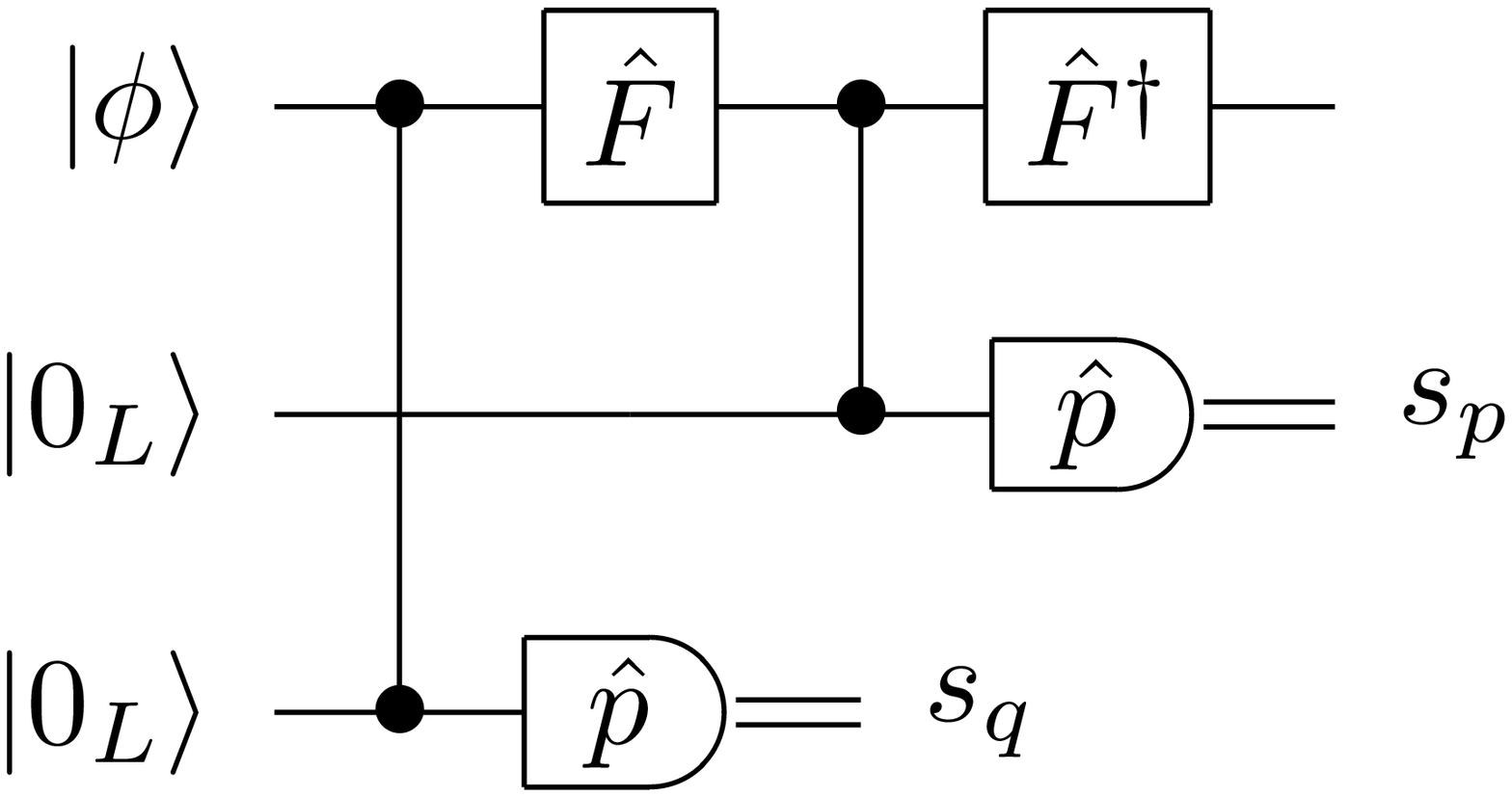}
\end{equation}
Up to Fourier transform and outcome-dependent displacements for correction, which are ubiquitous in measurement-based quantum computation anyway, this is quickly recognizable as a standard CV cluster-state computation with the following non-standard cluster:
\begin{align}
\label{cluster:EC}
    \Qcircuit[1.6em] @R=0.75em @C=0.75em {
    \node{0_\logical} \link{1}{0} & \node{0_\logical} \link{1}{0} \\
    \node{\phi} & \node{} \link{0}{-1} 
    }
\end{align}
As noted in the main text, the blank node represents a $\op p$-squeezed vacuum state (the input to a physical---and therefore imperfect---CV cluster state), and all links represent $\CZ$ gates of weight~$+1$. %
Measuring $\op p$ on each of the three marked nodes leaves a corrected version of $\ket \phi$ at the blank node.
\subsection{Evolution of the error matrix}

The error matrix~$\mat \eta$ for the GKP state can be thought of as the covariance matrix of a random Gaussian displacement (in momentum and position) that is applied to an otherwise perfect codeword state. If we write the column vector of Heisenberg-picture quadrature operators associated with the \emph{ideal} GKP state in question as
\begin{align}
\label{eq:xideal}
	\opvec x_\ideal = (\op q_{1,\ideal}, \dotsc, \op q_{N,\ideal}, \op p_{1,\ideal}, \dotsc, \op p_{N,\ideal})^\tp\,,
\end{align}
then the Heisenberg-picture quadrature operators associated with the physical approximation in question can be written as $\opvec x = \opvec x_\ideal + \vec y$, where $\vec y$ is a (classically) random displacement that is completely uncorrelated with the actual quantum information encoded in $\opvec x_\ideal$. To take an expectation value with respect to the quadratures~$\opvec x$ of the physical state, one may take the expectation value of the displaced ideal quadratures, $\opvec x_\ideal + \vec y$, and then average over the displacement~$\vec y$, which is Gaussian distributed with mean~0 and covariance matrix $\avg{ \vec y \vec y^\tp} = \mat \eta$. (Recall from the main text that the overall Gaussian envelope governing the spikes' heights is irrelevant if the error matrix is the same for each spike. The envelope will become important, however, when considering magic-state distillation, as discussed in Appendix~\ref{app:magic}.)

Since Gaussian unitary evolution corresponds to linear---and specifically, symplectic---evolution on phase space in the Heisenberg picture~\cite{Menicucci2011}, we can model this as
\begin{align}
	\opvec x' = \op U^\dag \opvec x \op U = \mat S \opvec x = \mat S \opvec x_\ideal + \mat S \vec y\,,
\end{align}
where $\mat S$ is a symplectic matrix representing the evolution. Notice that this has the form of evolving the ideal state and the random shift each by~$\mat S$:
\begin{align}
	\opvec x_\ideal &\mapsto \opvec x_\ideal' = \mat S \opvec x_\ideal\,, \\
	\vec y &\mapsto \vec y' = \mat S \vec y\,,
\end{align}
and thus $\opvec x' = \opvec x_\ideal' + \opvec y'$. The covariance matrix~$\mat \eta$ for the Gaussian random shift has now evolved to
\begin{align}
	\mat \eta \mapsto \mat \eta' = \avg{\vec y' \vec y'^\tp} = \mat S \avg{\vec y \vec y^\tp} \mat S^\tp = \mat S \mat \eta \mat S^\tp\,.
\end{align}

\subsection{Single-mode gates}

Given a measurement vector~$\measvec^{(G)}$ that implements gate $\op G$ on Cluster~\eqref{eq:singleGaussoncluster} in the main text, let us label the initial error matrix (i.e., the covariance matrix for the random shift in the initial state) by~$\mat \eta_0$. This error matrix is that for the state~$\phi$ indicated at node~1 in Cluster~\eqref{eq:singleGaussoncluster} in the main text. Although this localization of quantum information information in the cluster (implied by writing $\phi$ at that node) is not literally true since the information is, in fact, distributed throughout the cluster, it is common practice when discussing cluster-state computation to consider the logical information as being ``located at'' one particular node and then teleported (with a transformation applied) to an adjacent node via measurement. Such a description also has a precise mathematical interpretation~\cite{Nielsen2006,Alexander:MGl69zxv}, and as such we are safe to think in this picture.

Given a quantum state with error matrix $\mat \eta_{j-1}$ located at node~$j$, after the measurement of $\op p + \meas_j \op q$ on node~$j$, the symplectic matrix
\begin{align}
	\mat S_j &= \mat F \mat P(\meas_j) = \begin{pmatrix} 0 & -1 \\ 1 & 0 \end{pmatrix}
	\begin{pmatrix} 1 & 0 \\ \meas_j & 1 \end{pmatrix} = \begin{pmatrix} -\meas_j & -1 \\ 1 & 0 \end{pmatrix}
\end{align}
is applied to $\mat \eta_{j-1}$%
. This is accompanied by additional noise of variance~$\epsilon$ added in the $\op p$ quadrature after the applied gate, where $\epsilon$ is the variance of the initial $\op p$-squeezed states used to create the CV cluster state~\cite{Gu2009,Alexander:MGl69zxv}:
\begin{align}
	\mat \eta_{j-1} \mapsto \mat \eta_j = \mat S_j \mat \eta_{j-1} \mat S_j^\tp +
	\begin{pmatrix}
		0	&	0	\\
		0	&	\epsilon
	\end{pmatrix}
	\,.
\end{align}
The resulting $\mat \eta_j$ is now the error matrix for the new state located at node~$j+1$. After measurement \#3, there is a round of error correction. Assuming pure ancillas with symmetric noise of variance~$\delta$ in each quadrature, GKP error correction as implemented in Circuit~\eqref{circ:qcorrect} \emph{replaces} the original noise in $\op q$ with fixed and uncorrelated noise of variance~$\delta$ and \emph{adds} noise of variance~$\delta$ to the $\op p$ quadrature~\cite{Glancy2006}. This can be modeled as
\begin{align}
	\mat \eta_j 
	\mapsto
	\mat \eta_{j,\corr}
	\coloneqq
	\begin{pmatrix}
		0	&	0	\\
		0	&	1
	\end{pmatrix}
	\mat \eta
	\begin{pmatrix}
		0	&	0	\\
		0	&	1
	\end{pmatrix}
	+ \delta \mat \id
	\,,
\end{align}
where the subscript ($j,\corr$) stands for the corrected version of $\mat \eta_j$, and $\mat \id$ is the ${2\times 2}$ identity matrix. A final $\op p$ measurement on node~4 evolves the covariance matrix to
\begin{align}
	\mat \eta_{3,\corr} \mapsto \mat \eta_4 = \mat S_4 \mat \eta_{3,\corr} \mat S_4^\tp + 
	\begin{pmatrix}
		0	&	0	\\
		0	&	\epsilon
	\end{pmatrix}
	\,,
\end{align}
which then undergoes a final round of error correction analogous to the previous one:
\begin{align}
	\mat \eta_4 \mapsto \mat \eta_{4,\corr}\,.
\end{align}
The covariance matrix~$\mat \eta_{4,\corr}$ is the final error matrix in the state after all error correction is performed, and it becomes the new $\mat \eta_0$ for the next gate. This evolution is shown for the single-mode gates~$\op \id$, $\op P$, and $\op F$ in the first three columns of Table~\ref{tab:noiseevol}. Notice that $\mat \eta_0 = \mat \eta_{4,\corr}$ in each case. This is the reason for choosing $\mat \eta_0$ with asymmetric noise---it is the natural choice when considering sequential application of gates.

To figure out the probability that the gate undergoes a logical (i.e., qubit-level) error, we need the probability that either (or both) of the corrections at step~3 and~4 fail. Error correction proceeds by nondestructively measuring $\op q \mod \sqrt \pi$ using an ancilla that itself has an additional random shift error of variance~$\delta$ in each quadrature. This measurement effectively projects the error in~$\op q$ into a definite shift error~$\Delta q$, where the amount of the shift is known to accuracy~$\delta$. If $\abs{\Delta q} > \sqrt \pi/2$, then due to the modular arithmetic with range $[-\sqrt \pi/2, \sqrt \pi/2)$, it will be detected as being a shift of $\Delta q \pm \sqrt \pi$, resulting in a logical error at the qubit level. Since the random shift~$\Delta q$ is itself distributed according to a Gaussian with variance~$\eta_{j,qq}$, and since the accuracy to which this shift is known is also a Gaussian with variance~$\delta$, the total variance used for estimating the probability of a logical error in the correction at step~$j$ is
\begin{align}
	\sigma^2_{\err,j} = \eta_{j,qq}+\delta\,.
\end{align}
These variances are listed near the bottom of Table~\ref{tab:noiseevol}.

The probability that the correction at step~$j$ \emph{succeeds} is just the portion of a normalized Gaussian of variance~$\sigma^2_{\err,j}$ that lies between $\sqrt \pi/2$ and $-\sqrt \pi/2$:
\begin{align}
	p_{\success,j} &= \frac{1}{\sqrt{2 \pi \sigma^2_{\err,j}}} \int_{-\frac {\sqrt \pi} {2}}^{\frac {\sqrt \pi} {2}} dx\, \exp \left( - \frac {x^2} {2\sigma^2_{\err,j}} \right) \nonumber \\
	&= \erf \left( \frac {\sqrt \pi} {2\sqrt 2 \sigma_{\err,j}} \right)
	\,.
\end{align}
If we assume, as is done in the main text, that $\delta = {\epsilon \eqqcolon \sigma^2}$, then in all cases considered here, $\sigma^2_{\err,j} = n_j \sigma^2$ for some positive integer~$n_j$, and
\begin{align}
\label{eq:psuccess}
	p_{\success,j} &= \erf \left( \frac {\sqrt \pi} {2\sqrt {2n_j} \sigma} \right)
	\,.
\end{align}
For the gate to be free of error, both of the corrections at steps~3 and~4 must succeed. Therefore, the probability that the gate experiences a logical error is just one minus this:
\begin{align}
	p_\err = 1 - (p_{\success,3}) (p_{\success,4})\,.
\end{align}

\subsection{Two-mode gate (\texorpdfstring{$\CZ$}{CZ})}
\label{subapp:CZ}

The calculation for the probability of error in the $\CZ$ gate (implemented by $\op p$ measurements on Cluster~\eqref{eq:CVCZoncluster} in the main text) proceeds similarly to that for the single-mode gates described above. The main differences are (a)~that the error matrix~$\mat \eta$ is now ${4\times 4}$ to account for possible correlations between the error in each of the two modes and (b)~that the initial $\CZ$ gate introduces additional noise, thereby modifying $\mat \eta_0 \mapsto \mat \eta_0'$ before continuing with the usual propagation along the two rails.

To address~(a), us first set up the formalism to describe the error matrix for a two-mode state. We will use the convention that a ${4\times 4}$ covariance matrix is divided into four blocks as follows. Using the ordering described in Eq.~\eqref{eq:xideal}, we can see that the upper-left block contains the $q$-$q$ covariance matrix, and the bottom-right block contains the $p$-$p$ correlations, while the other two describe $q$-$p$ correlations. Given a two-mode quantum state with error matrix $\mat \eta_{j-1}$ located at nodes~$j$ in the top and bottom rails of Cluster~\eqref{eq:CVCZoncluster} in the main text, after the measurement of $\op q$ on top and bottom nodes~$j$, the error matrix evolves as
\begin{align}
	\mat \eta_{j-1} \mapsto \mat \eta_j =  \bar{\mat F} \mat \eta_{j-1} \bar{\mat F}^\tp +
	\begin{pmatrix}
		0 & 0 & 0 & 0 \\
		0 & 0 & 0 & 0 \\
		0 & 0 & \epsilon & 0 \\
		0 & 0 & 0 & \epsilon
	\end{pmatrix}
	\,,
\end{align}
where 
\begin{align}
	\bar{\mat F} &=
	\begin{pmatrix}
		0 & 0 & -1 & 0 \\
		0 & 0 & 0 & -1 \\
		1 & 0 & 0 & 0 \\
		0 & 1 & 0 & 0
	\end{pmatrix}
\end{align}
is the symplectic representation of the Fourier transform on two modes. The resulting $\mat \eta_j$ is now the error matrix for the new state located at nodes~$j+1$. Error correction at step~$j$ ($j = 3,4)$ now consists of two independent corrections but can be modeled as a single operation by the map
\begin{align}
	\mat \eta_j &\mapsto \mat \eta_{j,\corr} \coloneqq
	\begin{pmatrix}
		0 & 0 & 0 & 0 \\
		0 & 0 & 0 & 0 \\
		0 & 0 & 1 & 0 \\
		0 & 0 & 0 & 1
	\end{pmatrix}
	\mat \eta_j
	\begin{pmatrix}
		0 & 0 & 0 & 0 \\
		0 & 0 & 0 & 0 \\
		0 & 0 & 1 & 0 \\
		0 & 0 & 0 & 1
	\end{pmatrix}
	+ \delta \mat \id\,,
\end{align}
where $\mat \id$ now represents the ${4\times 4}$ identity matrix.

\begin{table*}[tp]
\ra{1}
\begin{tabular}{||c||c|c|c|c||}
\toprule
	& $\op \id$	& $\op P$		& $\op F$		& $\CZ[-1]$
\\
\toprule
	$\mat \eta_0$
&
	$\begin{pmatrix}
		\delta	&	0	\\
		0		&	2\delta + \epsilon
	\end{pmatrix}$
&
	$\begin{pmatrix}
		\delta	&	0	\\
		0		&	2\delta + \epsilon
	\end{pmatrix}$
&
	$\begin{pmatrix}
		\delta	&	0	\\
		0		&	2\delta + \epsilon
	\end{pmatrix}$
&
	$\left(\begin{smallmatrix}
		\delta	&	0		&	0				&	0	\\
		0		&	\delta	&	0				&	0	\\
		0		&	0		&	2\delta + \epsilon	&	0	\\
		0		&	0		&	0				&	2\delta + \epsilon
	\end{smallmatrix}\right)$
\\[3ex]
	$\mat \eta_0'$
&	--
&	--
&	--
&
	$\left(\begin{smallmatrix}
		\delta	&	0		&	0				&	-\delta	\\
		0		&	\delta	&	-\delta			&	0	\\
		0		&	-\delta	&	3\delta+ 2\epsilon	&	0	\\
		-\delta	&	0		&	0				&	3\delta + 2\epsilon
	\end{smallmatrix}\right)$
\\[3ex]
	$\mat \eta_1$
&
	$\begin{pmatrix}
		2\delta + \epsilon	&	0	\\
		0				&	\delta + \epsilon
	\end{pmatrix}$
&
	$\begin{pmatrix}
		3\delta + \epsilon	&	-\delta	\\
		-\delta			&	\delta + \epsilon
	\end{pmatrix}$
&
	$\begin{pmatrix}
		3\delta + \epsilon	&	-\delta	\\
		-\delta			&	\delta + \epsilon
	\end{pmatrix}$
&
	$\left(\begin{smallmatrix}
		3\delta+ 2\epsilon	&	0			&	0					&	\delta	\\
		0				&	3\delta+ 2\epsilon	&	\delta			&	0	\\
		0				&	\delta			&	\delta+ \epsilon	&	0	\\
		\delta			&	0				&	0				&	\delta + \epsilon
	\end{smallmatrix}\right)$
\\[3ex]
	$\mat \eta_2$
&
	$\begin{pmatrix}
		\delta + \epsilon	&	0	\\
		0				&	2\delta + 2\epsilon
	\end{pmatrix}$
&
	$\begin{pmatrix}
		\delta + \epsilon	&	\delta	\\
		\delta			&	3\delta + 2\epsilon
	\end{pmatrix}$
&
	$\begin{pmatrix}
		2\delta + 2\epsilon	&	-2\delta - \epsilon	\\
		-2\delta - \epsilon	&	3\delta + 2\epsilon
	\end{pmatrix}$
&
	$\left(\begin{smallmatrix}
		\delta+ \epsilon	&	0			&	0				&	-\delta	\\
		0			&	\delta+ \epsilon	&	-\delta			&	0	\\
		0			&	-\delta		&	3\delta+ 3\epsilon	&	0	\\
		-\delta		&	0			&	0				&	3\delta + 3\epsilon
	\end{smallmatrix}\right)$
\\[3ex]
	$\mat \eta_3$
&
	$\begin{pmatrix}
		2\delta + 2\epsilon	&	0	\\
		0				&	\delta + 2\epsilon
	\end{pmatrix}$
&
	$\begin{pmatrix}
		3\delta + 2\epsilon	&	-\delta	\\
		-\delta			&	\delta + 2\epsilon
	\end{pmatrix}$
&
	$\begin{pmatrix}
		\delta + 2\epsilon	&	-\epsilon	\\
		-\epsilon			&	2\delta + 3\epsilon
	\end{pmatrix}$
&
	$\left(\begin{smallmatrix}
		3\delta+ 3\epsilon	&	0			&	0					&	\delta	\\
		0				&	3\delta+ 3\epsilon	&	\delta			&	0	\\
		0				&	\delta			&	\delta+ 2\epsilon	&	0	\\
		\delta			&	0				&	0				&	\delta + 2\epsilon
	\end{smallmatrix}\right)$
\\[3ex]
	$\mat \eta_{3,\corr}$
&
	$\begin{pmatrix}
		\delta	&	0	\\
		0		&	2\delta + 2\epsilon
	\end{pmatrix}$
&
	$\begin{pmatrix}
		\delta	&	0	\\
		0		&	2\delta + 2\epsilon
	\end{pmatrix}$
&
	$\begin{pmatrix}
		\delta	&	0	\\
		0		&	3\delta + 3\epsilon
	\end{pmatrix}$
&
	$\left(\begin{smallmatrix}
		\delta	&	0		&	0				&	0	\\
		0		&	\delta	&	0				&	0	\\
		0		&	0		&	2\delta+ 2\epsilon	&	0	\\
		0		&	0		&	0				&	2\delta + 2\epsilon
	\end{smallmatrix}\right)$
\\[3ex]
	$\mat \eta_{4}$
&
	$\begin{pmatrix}
		2\delta + 2\epsilon	&	0	\\
		0				&	\delta + \epsilon
	\end{pmatrix}$
&
	$\begin{pmatrix}
		2\delta + 2\epsilon	&	0	\\
		0				&	\delta + \epsilon
	\end{pmatrix}$
&
	$\begin{pmatrix}
		3\delta + 3\epsilon	&	0	\\
		0				&	\delta + \epsilon
	\end{pmatrix}$
&
	$\left(\begin{smallmatrix}
		2\delta+ 2\epsilon	&	0				&	0				&	0	\\
		0				&	2\delta+ 2\epsilon	&	0				&	0	\\
		0				&	0				&	\delta+ \epsilon		&	0	\\
		0				&	0				&	0				&	\delta + \epsilon
	\end{smallmatrix}\right)$
\\[3ex]
	$\mat \eta_{4,\corr}$
&
	$\begin{pmatrix}
		\delta	&	0	\\
		0		&	2\delta + \epsilon
	\end{pmatrix}$
&
	$\begin{pmatrix}
		\delta	&	0	\\
		0		&	2\delta + \epsilon
	\end{pmatrix}$
&
	$\begin{pmatrix}
		\delta	&	0	\\
		0		&	2\delta + \epsilon
	\end{pmatrix}$
&
	$\left(\begin{smallmatrix}
		\delta	&	0		&	0				&	0	\\
		0		&	\delta	&	0				&	0	\\
		0		&	0		&	2\delta+ \epsilon	&	0	\\
		0		&	0		&	0				&	2\delta + \epsilon
	\end{smallmatrix}\right)$
\\[3ex]
\colrule
	$\sigma^2_{\err,3}$
&
	$3\delta + 2\epsilon$	& $4\delta + 2\epsilon$	& $2\delta + 2\epsilon$	& $4\delta + 3\epsilon$ (top rail)
\\
&						&					&					& $4\delta + 3\epsilon$ (bottom rail)
\\
\colrule
	$\sigma^2_{\err,4}$
&
	$3\delta + 2\epsilon$	& $3\delta + 2\epsilon$	& $4\delta + 3\epsilon$	& $3\delta + 2\epsilon$ (top rail)
\\
&						&					&					& $3\delta + 2\epsilon$ (bottom rail)
\\
\botrule
\end{tabular}
\caption{\label{tab:noiseevol}Evolution of the noise matrix~$\mat \eta$ under the single-mode Gaussian operations~$\op \id$, $\op P$, and $\op F$ and also under the two-mode Gaussian operation $\CZ[-1]$. The single-mode gates are implemented using the measurement vectors $\measvec^{(I)}$, $\measvec^{(P)}$, and $\measvec^{(F)}$, respectively, on Cluster~\eqref{eq:singleGaussoncluster} in the main text. The $\CZ[-1]$ gate is implemented using all $\op p$ measurements on Cluster~\eqref{eq:CVCZoncluster} in the main text. $\sigma^2_{\err,j}$ is related to the probability of error in the correction (higher $\to$ error is more likely). See text for further details.
}
\end{table*}

To address~(b), consider the vertical portion of Cluster~\eqref{eq:CVCZoncluster} in the main text, now turned on its side:
\begin{align}
\label{eq:CZreduce}
    \Qcircuit[1.6em] @R=0.75em @C=0.75em {
    \node{\phi} {} & \node{} \link{0}{-1} & \node{} \link{0}{-1} & \node{\psi} \link{0}{-1}  
    \\
    {\vphantom{b}a} & {b} & {\vphantom{b}c} & {d}
    }
    \,,
\end{align}
where the weight of each link is~$+1$, and consider any input pure state~$\op \rho_0 = \outprod\phi \phi \otimes \outprod\psi \psi$. Measuring $\op p$ on nodes~$b$ and~$c$, performing the required outcome-dependent displacements on nodes~$a$ and $d$, and averaging over the measurement outcomes~\cite{Gu2009,Alexander:MGl69zxv} results in the following cluster on the remaining nodes:
\begin{align}
\label{eq:CZreducefinal}
    \Qcircuit[1.6em] @R=0.75em @C=0.75em {
    \node{\phi'} {} & \displaystyle{\qquad {-1} \atop {}} & & \node{\psi'} \link{0}{-3}  
    \\
    {\vphantom{b}a} & & & {d}
    }
    \,,
\end{align}
where the resulting weight, as indicated, is now~$-1$ and where the primes on the states indicates that each mode has separately undergone a random Gaussian shift in $\op p$ with variance~$\epsilon$. By linearity, then, the average map for \emph{any} input state~$\op \rho$ under these measurements is therefore
\begin{align}
	\op \rho &\mapsto \cG \bigl( \CZ[-1] \op \rho \CZ[-1]^\dag \bigr) \nonumber \\
	&= \CZ[-1] \cG (\op \rho) \CZ[-1]^\dag\,,
\end{align}
where $\cG$ represents the addition of uncorrelated noise of variance~$\epsilon$ to the $\op p$ quadrature of each mode. Notice that this noise application commutes with the $\CZ$ gate. Also notice that the $\CZ$ gate has a weight of~$-1$. Ordinarily this would be important, but because the encoded controlled-$Z$ gate can be represented by either~$\CZ[\pm 1]$, we don't need to worry about this change of weight when considering its action on encoded states.

We must consider the action of this noisy $\CZ$ operation on the initial error matrix. Given a ${4\times 4}$ error matrix~$\mat \eta_0$ as the output of a previous computation, under the $\CZ[-1]$ and noise~$\cG$, this error matrix becomes
\begin{align}
	\mat \eta_0 \mapsto \mat \eta_0' = \mat{C_Z}[-1] \mat \eta_0 \mat{C_Z}[-1]^\tp +
		\begin{pmatrix}
		0 & 0 & 0 & 0 \\
		0 & 0 & 0 & 0 \\
		0 & 0 & \epsilon & 0 \\
		0 & 0 & 0 & \epsilon
	\end{pmatrix}
	\,,
\end{align}
where 
\begin{align}
	\mat {C_Z}[-1] &=
	\begin{pmatrix}
		1 & 0 & 0 & 0 \\
		0 & 1 & 0 & 0 \\
		0 & -1 & 1 & 0 \\
		-1 & 0 & 0 & 1
	\end{pmatrix}
\end{align}
is the symplectic representation of $\CZ[-1]$. The evolution implemented by the subsequent $\op p$ measurements will therefore begin with~$\mat \eta_0'$ instead of $\mat \eta_0$ at the first step. The evolution of the noise matrix under the $\CZ[-1]$ gate implemented by Cluster~\eqref{eq:CVCZoncluster} in the main text is shown in the fourth column of Table~\ref{tab:noiseevol}. This column is the only one to have an entry for $\mat \eta_0'$ since this error matrix (after the $\CZ$ gate but before teleportation to the next node pair) is only defined for the $\CZ$ gate. Notice that, once again, $\mat \eta_0 = \mat \eta_{4,\corr}$ and that this is just two copies of the $\mat \eta_0$ used in the single-mode case above, guaranteeing that correlated errors cannot build up. Notice that there are now a total of four independent instances of error correction, consisting of two at step~3 and two more at step~4. Each corresponds to an error probability easily generalized from the single-mode case using
\begin{align}
	\sigma^2_{\err,j,\toplab} &= \eta_{j,{q_1}{q_1}}+\delta & &\text{(top rail)}\,, \nonumber \\
	\sigma^2_{\err,j,\botlab} &= \eta_{j,{q_2}{q_2}}+\delta & &\text{(bottom rail)}\,.
\end{align}
These variances are listed near the bottom of Table~\ref{tab:noiseevol} in the fourth column. As expected by the symmetry of the gate, $\sigma^2_{\err,j,\toplab} = \sigma^2_{\err,j,\botlab} = n_j \sigma^2$, where we have defined $\sigma^2 \coloneqq \delta = \epsilon$ as before. We can therefore write the probability of an error as one minus the probability that all corrections succeed:
\begin{align}
\label{eq:perrCZ}
	p_\err = 1 - (p_{\success,3})^2 (p_{\success,4})^2\,,
\end{align}
where the squares are due to there being two corrections at each step. From Table~\ref{tab:noiseevol}, $n_3 = 7$, and $n_4 = 5$. Plugging these into Eq.~\eqref{eq:perrCZ} and using Eq.~\eqref{eq:psuccess} gives the error probability quoted in the main text [Eq.~(3)]. Table~\ref{tab:noiseevol} confirms that the $\CZ$ gate has the highest error probability of any of the gates considered. %

\section{Magic-state distillation calculations}
\label{app:magic}

\subsection{Method}
To prepare a Hadamard eigenstate, we use the following procedre. We begin with Cluster~\eqref{eq:CVCZoncluster} of the main text with $\ket \phi \otimes \ket \psi = \ket {+_\logical} \otimes \ket {+_\logical}$. (Considering Cluster~\eqref{eq:CVCZoncluster} as part of a regular lattice supplemented with ancillas at appropriate intervals, there will be an ancilla above the first node on each rail, which can be measured in~$\op p$ to prepare this initial state.) Implementing the $\CZ$ gate with $\op p$ measurements results in an encoded (and error-corrected) Bell pair at the output nodes---specifically, an encoded two-qubit cluster state, $\frac {1} {\sqrt 2} \ket {0_\logical}\ket{+_\logical} + \frac {1} {\sqrt 2} \ket {1_\logical}\ket{-_\logical}$. We then flip a fair coin and perform an encoded Hadamard on the bottom-rail qubit if it comes up heads. On average, this random operation decoheres the Bell pair into the Hadamard eigenbasis: $\frac 1 2 \outprod {+H_\logical} {+H_\logical}^{\otimes 2} + \frac 1 2 \outprod {-H_\logical} {-H_\logical}^{\otimes 2}$, which can be interpreted as classical ignorance of whether the state is actually $\ket {+ H_\logical}^{\otimes 2}$ or $\ket {- H_\logical}^{\otimes 2}$ (with equal probability). Properly identifying the state of the top mode through destructive photon counting will therefore reveal which state exists in the bottom mode.

Noise will corrupt this measurement. GKP point out that this procedure works best with uncorrelated, isotropic noise~\cite{Gottesman2001}, but the GKP error correction we have been using produces  noise that is asymmetric in $\op q$ and $\op p$---i.e., $\mat \eta = \left( \begin{smallmatrix} \delta & 0 \\ 0 & 2\delta + \epsilon \end{smallmatrix}\right)$---even as the noise inherent to CV cluster-state computation is symmetric on average~\cite{Gu2009,Alexander:MGl69zxv}. We can fix this by intentionally blurring the $\op q$ quadrature with a (classically) random Gaussin shift, which, by assumption, costs nothing to employ. Then, we count photons. An odd outcome forces us to start over. Given an even outcome, $\varepsilon$ is the probability that it reveals the wrong state. %

This discussion of magic-state distillation mentioned the assumption that the ancillas are pure. In what follows, we explain why this assumption is important, and we also describe the details of calculating the error probability and success probability, both performed in the Wigner picture.

\subsection{Importance of the Gaussian envelope}

Throughout the discussion of Clifford gates, we have ignored entirely the question of the size of the Gaussian envelope used to regulate the height of the spikes in the Wigner-function representation of the encoded states. This is because, as mentioned in the main text and in Appendix~\ref{app:clifford}, this envelope is irrelevant when considering GKP error correction: as long as each spike has the same error matrix, it makes no difference how big one spike is relative to any other. It is of \emph{crucial} importance, however, when considering photon counting because an envelope that is ``too big'' will raise the error rate. This can be understood intuitively by considering that the Wigner functions for photon-number eigenstates, as the photon number increases, (a)~have support further from the origin and (b)~get more and more oscillatory. Thus, Gaussian blurring of a centrally located spike will be far less likely to result in the wrong number of photons being counted than will the same blurring of a spike very far from the origin. Thus, for photon counting, smaller envelopes are better. But we know that the envelope cannot be too small, for if it were small enough to only contain just a single narrow spike, this would violate the uncertainty principle, which is impossible.

So far, we have been thinking about the calculation in terms of the Wigner function since the error matrix fits naturally into this context as the covariance matrix of a Gaussian blurring function applied to the Wigner-picture lattice of $\pm$~$\delta$-functions representing GKP codeword states~\cite{Gottesman2001}. But let us shift gears for a moment and consider the (normalized) wavefunction for an arbitrary pure state: $\psi(s) = \inprodsubsub s \psi q {}$. A Gaussian envelope with variance~$\xi^2$ in the position basis multiplying that state---while \emph{ignoring normalization}---results in
\begin{align}
	\psi(s) &\mapsto \exp \left( \frac {-s^2} {2\xi^2} \right) \psi(s)\,,
\end{align}
which can be written
\begin{align}
	\inprodsubsub s \psi q {} &\mapsto e^{-s^2/2\xi^2} \inprodsubsub s \psi q {} = \qbra s e^{-\op q^2/2\xi^2} \ket \psi\,.
\end{align}
Therefore, the nonunitary operation that applies an envelope in position is $e^{-\op q^2/2\xi^2}$ (up to normalization), which can also be interpreted equivalently as a convolution of the momentum-space wavefunction by a Gaussian of variance~$1/\xi^2$. Similarly, to multiply the momentum-space wavefunction of a state by an envelope with variance~$\xi^2$ (again, ignoring the required renormalization) or, equivalently, to convolve the position-space wavefunction by a Gaussian with variance~$1/\xi^2$, one can simply apply the nonunitary operator~$e^{-\op p^2/2\xi^2}$. For large envelopes ($\xi^2 \gg 1$), these operators approximately commute, and we can combine them into a single nonunitary operator that does both envelopes at the same time:
\begin{align}
	e^{-\op q^2/2\xi^2} e^{-\op p^2/2\xi^2} &= \exp \left[ -\frac {\left(2\op a^\dag \op a + 1\right)} {2\xi^2} + O\left( \frac {1} {\xi^4} \right) \right] \nonumber \\
	&\simeq e^{-1/2\xi^2} \exp \left( -\frac {\op a^\dag \op a} {\xi^2} \right)\,.
\end{align}
Ignoring the constant exponential (since the state has to be renormalized anyway), the remaining operator
\begin{align}
	\op \cool = \exp \left( -\frac {\op a^\dag \op a} {\xi^2} \right)
\end{align}
can be interpreted as a ``cooling'' operation since it reduces the temperature of thermal states (up to normalization). Writing an arbitrary pure state~$\ket \psi$ in its number-state decomposition,
\begin{align}
	\ket \psi &= \sum_{n\in\wholenums} c_n\ket n\,,
\intertext{where $\wholenums$ is the set of nonnegative integers, we have}
	\op \cool \ket \psi &= \sum_{n \in \wholenums} e^{-n/\xi^2} c_n \ket n\,.
\end{align}
The action of $\op \cool$ is simply to damp out the high-number components of the state vector (resulting in an unnormalized vector). Because $\op \cool$ is diagonal in the number basis, \emph{noise of this form cannot cause photon-counting errors}.

Let us consider what noise of this form looks like in the Wigner picture. Obviously, we can consider writing the state in the number basis and damping high-number terms, but the Wigner function is more suited to a phase-space analysis. As such, let us continue to assume that $\xi^2 \gg 1$. The operator~$e^{-\op q^2/2\xi^2}$ multiplies the position-space wavefunction by a Gaussian envelope with variance~$\xi^2$, which corresponds to two actions on the associated Wigner function. First, it applies a Gaussian envelope in position to the Wigner function. This Wigner-picture envelope has variance~$\xi^2/2$ due to the fact that the wavefunction envelope gets squared when considering its action on the Wigner function. Second, it \emph{simultaneously} convolves the Wigner function in momentum with a Gaussian of variance~$1/2\xi^2$. Again, the fact that the momentum-space wavefunction gets squared when calculating probabilities is responsible for the factor of two in this variance. The operator~$e^{-\op p^2/2\xi^2}$ behaves analogously with position and momentum exchanced.  When $\xi^2 \gg 1$, the application of a Wigner-picture Gaussian envelope and a Wigner-picture convolution (in the same quadrature) approximately commute, and we can model the action of $\op \cool$ as
\begin{align}
	W (q,p) \mapsto W'(q,p) \simeq e^{-(q^2 + p^2)/\xi^2} [W * G_{1/2\xi^2}] (q,p)\,,
\end{align}
where $*$ indicates convolution with respect to both arguments, and
\begin{align}
\label{eq:Gaussnorm}
	G_{1/2\xi^2} (q,p) = \frac {\xi^2} {\pi} e^{-\xi^2(q^2+p^2)}
\end{align}
is a normalized isotropic Gaussian with variance~$1/2\xi^2$ in each quadrature. Also keep in mind that $W'(q,p)$ is not normalized. We do not have to normalize it now if we promise to take this into account properly when taking expectation values.

\subsection{Pure ancilla states}

Notice that ``cooling'' a state with $\op \cool$ produces another pure state and also does not cause any photon-counting error. This is the motivation for assuming that the ancilla states are pure: we can write them as having noise of the form generated by $\op \cool$. Specifically, an approximate ancilla state~$\ket{0_\logical}$ is related to an ideal ancilla state~$\ket{0_{\logical,\ideal}}$ by
\begin{align}
	\ket{0_\logical} \propto
	{e^{-2\delta \op a^\dag \op a} \ket{0_{\logical,\ideal}}} 
	\,,
\end{align}
where renormalization is implied. The parameter~$\xi^2$ is chosen to be $\xi^2 = 1/2\delta$ so that the variance of the Wigner-function blurring noise is~$\delta$, as we have been using. The corresponding Wigner-function envelope has variance~$1/4\delta$, which ensures that the state is pure. The assumption of pure ancillas limits the number of photons available in the first place, thereby reducing the chance that a high-number component of the state experiences Gaussian blurring, which would result in photon-counting error.

A pure ancilla will also reapply a properly sized Gaussian envelope at each stage of error correction. This new envelope in \emph{position} is responsible for the additional noise in \emph{momentum} added to the error matrix. The mean of the envelope, however, is the particular outcome~$s_q$ of the measurement made in Circuit~\ref{circ:qcorrect}, and this can be quite far from~0 if the ancilla has very low error. The prescribed correction after Circuit~\ref{circ:qcorrect} only shifts the state back by the minimum amount required to reenter the codespace [i.e., by $-(s_q \mod \sqrt\pi)$], but doing this assumes full translational invariance of the codespace, which is not actually the case since spikes located far from the origin correspond to much higher energies. While this procedure is fine for correcting Gaussian noise, this minimal shift causes problems if we want to do reliable photon counting. As such, we can produce a more central state by applying the additional correction
\begin{align}
\label{eq:Xrecenter}
	\op X\left(-\left\lfloor s_q \right\rceil_{2\sqrt\pi} \right)\,,
\end{align}
where $\lfloor x \rceil_{2\sqrt\pi}$ is the integer multiple of~$2\sqrt\pi$ that is closest to~$x$. This operation does not change the logical state. For large envelopes, $(\sqrt\pi)^2 \ll 1/2\delta$, and thus, variation in the center of the envelope is a small correction that can be ignored. By using this modified form of error correction, we can use the pure ancillas to produce a state that has a centered envelope with variance approximately equal to that of the pure ancillas.

Propagation through the cluster, on the other hand, is assumed to entail only blurring noise, with no additional Gaussian envelope on average~\cite{Gu2009,Alexander:MGl69zxv}. This means that states that result from the final error-correction step at any gate will have approximately an overall Wigner-picture envelope of variance~$1/4\delta$ in both quadratures (from the pure ancillas) but could have spikes that are ``fatter'' than they should be for that size envelope.

In particular, the procedure outlined in the main text (after the classical blurring in~$\op q$), results in a final error matrix of~$\mat \eta = \left( \begin{smallmatrix} 3\sigma^2 & 0 \\ 0 & 3\sigma^2 \end{smallmatrix} \right)$, where we have assumed, as usual, that $\delta = \epsilon = \sigma^2$. This means that the variance of the Wigner-function spikes is~$3\sigma^2$, while the envelope has a variance of~$1/4\sigma^2$, and thus, the Wigner-function spikes are \emph{three times} as ``fat'' as they should be in terms of their variance. This is responsible for the high error probability~$\varepsilon$ %
reported in the main text, as we will see shortly.

The probabilities calculated below and reported in the main text are within the distillation threshold~\cite{Reichardt:2005er}, but we could do better with either of the following modifications to the procedure. The first would be to modify the outcome-dependent corrections used in the cluster-state information processing since there is no reason to preserve the high-amplitude components of an encoded state---shifting all the spikes back toward the origin is just as good in terms of the encoded information (and better for photon counting). A shift of the form of Eq.~\eqref{eq:Xrecenter} could therefore be applied in addition to the ordinary outcome-dependent shift, which would similarly recenter the state's envelope after each step through the cluster. Second, one could imagine reducing the error probability at the cost of reduced success probability simply by discarding photon counts that are higher than some maximum number.

In what follows, we \emph{do not} make either of these modifications in order to show the existence of a squeezing threshold with a minimum of additional assumptions and also for calculational simplicity. If we wanted to optimize the distillation procedure, however, these are two possible ways to do so.

\subsection{Probability formulas}

Let
\begin{align}
	\op \pi_\pm &\coloneqq \ket {\pm H_{\logical,\ideal}} \bra {\pm H_{\logical,\ideal}}
\end{align}
be the projector onto the ideal ${\pm 1}$~eigenstate of the encoded Hadamard operator. Further, let $\cM$ be the superoperator representing the application of an isotropic Wigner-function Gaussian envelope of variance~$1/4\sigma^2$ in each quadrature and a Wigner-function blurring by an isotropic Gaussian of variance $3\sigma^2$. This is a physical operation because it can be modeled as cooling with $\op \cool$ (with $\xi^2 = 1/2\sigma^2$), which applies the desired envelope and then blurs the spikes by an isotropic Gaussian of variance~$\sigma^2$ in each quadrature, followed by additional blurring by an isotropic Gaussian of variance~$2\sigma^2$ in each quadrature. Notice that this operation produces an unnormalized state. Using this superoperator, the state we have prepared is therefore, after normalization,
\begin{align}
\label{eq:}
	\op \rho_\pm \coloneqq \frac {\cM(\op \pi_\pm)} {\tr \bigl[ \cM(\op \pi_\pm) \bigr]}\,,
\end{align}
with equal probability for~$\pm$. We want to count photons on this state, but all we care about is the outcome modulo~4. As such, we can coarse-grain the measurement operators by defining the projectors
\begin{align}
	\op \Pi_a \coloneqq \sum_{n \in 4\wholenums + a} \outprod n n\,,
\end{align}
which each project onto the subspace of states that have $a$~photons (modulo~4). Notice that $\sum_{a \in \integers_4} \op \Pi_a = \op \id$.
While $\op\rho_\pm$ is normalized, it will not always be necessary to use it this way. Instead, we can work directly with $\cM(\op\pi_\pm)$ by defining
\begin{align}
\label{eq:Anorm}
	A[\cdot|\pm] \coloneqq \tr \left[\cM(\op \pi_\pm) \right] = \tr \left[\cM^* \bigl(\op \id \bigr) \op \pi_\pm \right]
\end{align}
and
\begin{align}
\label{eq:Aa}
	A[a|\pm] \coloneqq \tr \left[ \op\Pi_a \cM(\op \pi_\pm) \right] = \tr \left[ \cM^*\bigl( \op\Pi_a \bigr) \op \pi_\pm \right]\,,
\end{align}
where we have moved the noise to the measurement operator instead. The relevant probabilities are then given by
\begin{align}
\label{eq:PfromA}
	P[a|\pm] &= \frac {A[a|\pm]} {A[\cdot|\pm]}\,,
\end{align}
which represent the probability of obtaining outcome~$a$ when counting the number of photons (modulo~4), given that the state actually prepared was $\op\pi_\pm$, respectively. Since each initial state ($\pm$) is equally likely, the error probability, \emph{conditioned on obtaining an even number of photons}, is 
\begin{align}
\label{eq:perrdistill}
	\varepsilon %
	&= \frac {P[2|+] + P[0|-]} {P[0|+] + P[2|+] + P[0|-] + P[2|-]}\,.
\end{align}
The success probability (i.e., probability of getting an even outcome) in this case is
\begin{align}
\label{eq:psuccdistill}
	P[\text{even}] &= \frac 1 2 \Bigl(P[0|+] + P[2|+] + P[0|-] + P[2|-] \Bigr)\,.
\end{align}
All of these probabilities build on~$A[\cdot|\pm]$ and~$A[a|\pm]$. We will evaluate them in the Wigner picture.

\subsection{Wigner-picture probability calculation}

Because $\sigma^2 \ll 1$, the operations of blurring and applying the envelope approximately commute, and thus~$\cM = \cM^*$. %
Since applying applying a Gaussian envelope to a Wigner function is easy, all we need to explicitly calculate is the blurred version of the Wigner function representing~$\op \Pi_a$. We begin with some definitions. The number-state projector~$\outprod n n$ has the Wigner function~\cite{Cahill:1969jg,Cahill:1969it}
\begin{align}
	W_n(\vec r) &= \frac 1 \pi (-1)^n L_n(2r^2)e^{-r^2}\,,
\end{align}
where $\vec r = (q,p)^\tp$, and $r = \abs{\vec r}$. The formula for the trace of a product of two Hermitian operators, evaluated in the Wigner picture, is
\begin{align}
\label{eq:Wigneroverlap}
	\tr (\op A \op B) = 2\pi \int d^2 r W_A(\vec r) W_B(\vec r)\,,
\end{align}
where $W_A(\vec r)$ is the Wigner function associated with~$\op A$, and likewise for~$W_B(\vec r)$ and~$\op B$.

We would like to eventually calculate the Wigner functions for the modulo-4 projectors~$\op \Pi_a$. To this end, we first consider the following four (rather pathological) operators ($b \in \integers_4$):
\begin{align}
	\op \Phi_b \coloneqq \sum_{n \in \wholenums} i^{bn} \outprod n n\,,
\end{align}
where $i^{bn}$ should be interpreted as $e^{i b n \pi/2}$. Note that $\op \Phi_0 = \op \id$.  We can take linear combinations  of these operators to obtain
\begin{align}
	\op \Pi_a &= \frac 1 4 \sum_{b \in \integers_4} i^{-ab} \op \Phi_b 
	\,.
\end{align}
We would like, then, to find Wigner representations of $\op \Phi_a$. The following formal series will be useful:
\begin{align}
	\sum_{n\in \wholenums}  t^n L_n(x) = \frac {1} {1-t} \exp \left( \frac {-tx} {1-t} \right)%
	\,.
\end{align}
Since the definition of the Wigner function is linear in the operator being represented, we can write the Wigner functions for $\op \Phi_b$ formally as
\begin{align}
	W_{\Phi_b}(\vec r) &= \sum_{n \in \wholenums} i^{bn} W_n(\vec r) \nonumber \\
	&= \sum_{n \in \wholenums} i^{bn} \frac 1 \pi (-1)^n L_n(2r^2)e^{-r^2} \nonumber \\
	&= \frac {e^{-r^2}} \pi \sum_{n \in \wholenums} (-i^b)^n L_n(2r^2)\,.
\end{align}
Unfortunately, this sum will not converge in all cases. On physical grounds, we can instead consider the cooled version of these operators. 
Let us insert an exponential convergence factor into the above expressions (and later take the limit $\beta \to 0$):
\begin{align}
	W_{\Phi_b}(\vec r;\beta) &\coloneqq \sum_{n \in \wholenums} i^{bn} e^{-\beta n} W_n(\vec r) \nonumber \\
	&= \frac {e^{-r^2}} \pi \sum_{n \in \wholenums} (-i^b e^{-\beta})^n L_n(2r^2) \nonumber \\
	&= \frac {1} \pi \frac {1} {1+e^{-\beta+ib\pi/2}} \exp \left( -\frac {1-e^{-\beta+ib\pi/2}} {1+e^{-\beta+ib\pi/2}} r^2 \right)\,.
\end{align}
The case $b=2$ is problematic. The others present no problem, however, and for all cases, we redefine ${W_{\Phi_b}(\vec r) \coloneqq \lim_{\beta \to 0^+} W_{\Phi_b}(\vec r;\beta)}$. In this limit, we have
\begin{align}
	W_{\Phi_0}(\vec r) &= \frac {1} {2\pi}\,, \\
	W_{\Phi_1}(\vec r) &= \frac {1} {2\pi} (1-i)e^{i r^2}\,, \\
	W_{\Phi_2}(\vec r) &= \frac {1} {2\pi} \pi \delta^2(\vec r)\,, \\
	W_{\Phi_3}(\vec r) &= \frac {1} {2\pi} (1+i)e^{-i r^2}\,.
\end{align}
We can now write the Wigner functions of interest,
\begin{align}
	W_{\Pi_a}(\vec r) = \frac 1 4 \sum_{b \in \integers_4} i^{-ab} W_{\Phi_b}(\vec r)\,,
\end{align}
which become
\begin{align}
\label{eq:Pi0Wigner}
	W_{\Pi_0}(\vec r) &= \frac {1} {8\pi} \left[ 1 + \pi \delta^2(\vec r) + 2 \sin r^2 + 2 \cos r^2 \right]\,, \\
	W_{\Pi_1}(\vec r) &= \frac {1} {8\pi} \left[ 1 - \pi \delta^2(\vec r) + 2 \sin r^2 - 2 \cos r^2 \right]\,, \\ 
\label{eq:Pi2Wigner}
	W_{\Pi_2}(\vec r) &= \frac {1} {8\pi} \left[ 1 + \pi \delta^2(\vec r) - 2 \sin r^2 - 2 \cos r^2 \right]\,, \\ 
	W_{\Pi_3}(\vec r) &= \frac {1} {8\pi} \left[ 1 - \pi \delta^2(\vec r) - 2 \sin r^2 + 2 \cos r^2 \right]\,. 
\end{align}
Although we only need the even ones, it is instructive to see the pattern, so all four are included here. We can analytically convolve these functions with an isotropic Gaussian of variance~$\tau^2$,
\begin{align}
\label{eq:Gaussnorm}
	G_{\tau^2}(\vec r) \coloneqq \frac {1} {2\pi \tau^2} \exp \left( - \frac {r^2} {2\tau^2} \right)\,,
\end{align}
to obtain the following blurred versions (now keeping only the even ones):
\begin{widetext}%
\begin{align}
	W_{\Pi_0}(\vec r; \tau^2) &\coloneqq [W_{\Pi_0} * G_{\tau^2}](\vec r) \nonumber \\
	&= \frac {1} {8\pi} \left\{ 1 + \pi G_{\tau^2}(\vec r) + \frac{2 e^{-\frac{2 r^2 \tau ^2}{4 \tau ^4+1}}}{4 \tau ^4+1} \left[ \left(1-2 \tau ^2\right) \sin \left(\frac{r^2}{4 \tau ^4+1}\right)+\left(2 \tau ^2+1\right) \cos \left(\frac{r^2}{4 \tau ^4+1}\right)\right] \right\}\,, \\
	W_{\Pi_2}(\vec r; \tau^2) &\coloneqq [W_{\Pi_2} * G_{\tau^2}](\vec r) \nonumber \\
	&= \frac {1} {8\pi} \left\{ 1 + \pi G_{\tau^2}(\vec r) - \frac{2 e^{-\frac{2 r^2 \tau ^2}{4 \tau ^4+1}}}{4 \tau ^4+1} \left[ \left(1-2 \tau ^2\right) \sin \left(\frac{r^2}{4 \tau ^4+1}\right)+\left(2 \tau ^2+1\right) \cos \left(\frac{r^2}{4 \tau ^4+1}\right)\right] \right\}\,,
\end{align}%
\end{widetext}%
which can clearly be seen to reduce to Eqs.~\eqref{eq:Pi0Wigner} and~\eqref{eq:Pi2Wigner} in the limit~$\tau^2 \to 0$.

Now we need the Wigner functions for the (ideal) Hadamard-eigenstate projectors~$\op \pi_\pm$. We can build these up most easily by noting that
\begin{align}
\label{eq:Hassum}
	\op\pi_\pm = \frac {1} {\sqrt 2} \op X_\logical + \frac {1} {\sqrt 2} \op Z_\logical - \frac {\sqrt 2 - 1} {2} \op \id_\logical\,,
\end{align}
where $\op X_\logical$ and $\op X_\logical$ are encoded Pauli operators, and $\op \id_\logical$ is the projector onto the ideal codespace. Using the Wigner functions for the GKP basis states, Eq.~(28) of Ref.~\cite{Gottesman2001}, one can build up the Wigner functions for the encoded Pauli matrices and for the encoded identity and then take the linear combination specified in Eq.~\eqref{eq:Hassum} to obtain
\begin{align}
	W_{\pm H}(\vec r) = \sum_{s,t\in \integers} \delta \left(p-\frac{\sqrt{\pi } s}{2}\right) \delta \left(q-\frac{\sqrt{\pi } t}{2}\right) \lambda_j (t,s)
\end{align}
as the Wigner function corresponding to $\op\pi_\pm$, respectively. The Hadamard \emph{indicator function}~$\lambda$ is defined to be
\begin{align}
	\lambda_j(t,s) &=
	\begin{cases}
		\frac {1} {\sqrt 2} (-1)^{j+\frac{t}{2}} & \text{if $t$ even, $s$ odd;} \\
		\frac {1} {\sqrt 2} (-1)^{j+\frac{s}{2}} & \text{if $s$ even, $t$ odd;} \\
		1 & \text{if $t$ even, $s$ even;} \\
		0 & \text{if $t$ odd, $s$ odd,}
	\end{cases}
\end{align}
where $j=0,1$ corresponds to the state~$\op\pi_\pm$, respectively. Notice that these Wigner functions are not normalized, nor are they even normalizable. This is okay since we are assuming an unnormalized state anyway and normalizing within the calculation itself.

We are now ready to calculate Eqs.~\eqref{eq:Anorm} and~\eqref{eq:Aa} in the Wigner representation. Using Eq.~\eqref{eq:Wigneroverlap}, and recalling that $\op \Phi_0 = \op \id$ (which is invariant under blurring), we have
\begin{align}
	A[\cdot|\pm] &= \tr \left[\cM^* \bigl(\op \id \bigr) \op \pi_\pm \right] \nonumber \\
	&= 2\pi \int d^2 r\, G_{1/4\sigma^2}(\vec r) W_{\Phi_0}(\vec r) W_{\pm H}(\vec r) \nonumber \\
	&= \sum_{s,t\in \integers} G_{1/4\sigma^2}\left(\frac{\sqrt{\pi } s}{2}, \frac{\sqrt{\pi } t}{2}\right) \lambda_j (t,s)
	\,.
\end{align}
Similarly,
\begin{align}
	A[a|\pm] &= \tr \left[\cM^* \bigl(\op \Pi_a \bigr) \op \pi_\pm \right] \nonumber \\
	&= 2\pi \int d^2 r\, G_{1/4\sigma^2}(\vec r) W_{\Pi_a}(\vec r; 3\sigma^2) W_{\pm H}(\vec r) \nonumber \\
	&= 2\pi \sum_{s,t\in \integers} G_{1/4\sigma^2}\left(\frac{\sqrt{\pi } s}{2}, \frac{\sqrt{\pi } t}{2}\right) \nonumber \\
	&\qquad \times W_{\Pi_a}\left(\frac{\sqrt{\pi } s}{2}, \frac{\sqrt{\pi } t}{2}; 3\sigma^2\right) \lambda_j (t,s)
	\,.
\end{align}
These equations were summed numerically using \emph{Mathematica~9} and plugged into Eq.~\eqref{eq:PfromA} to get the relevant probabilities, which were then plugged into Eqs.~\eqref{eq:perrdistill} and~\eqref{eq:psuccdistill} and reported in the main text.

It is curious that the error probability~$\varepsilon$ varies so little over the different squeezing levels, and it is also curious that the success probability is almost exactly~$2/3$. One likely explanation is that some sort of analytic approximation to these equations can be made in the limit $\sigma^2 \to 0$ so that the only thing that becomes important is the product of blurring variance to envelope variance. When their product is~$1/4$, the error probability should go to zero. In the case above, this product is $3/4$, which corresponds to~$\varepsilon \simeq 12.6\%$. If we could design a protocol for which this product is $1/2$, the error probability reduces to $\varepsilon \simeq 5.6\%$, and the success probability jumps to~$3/4$. The fact that the success probability always seems to be a simple fraction also suggests a missing simplification. These questions are left to future work.